%% file: main.tex
\documentclass[12pt,onecolumn,prd,tightenlines,%
superscriptaddress,nofootinbib,%
preprintnumbers]{revtex4-2}
\usepackage[english]{babel}
\usepackage{graphicx}
\usepackage{bm}
\usepackage{slashed}
\usepackage{xspace}

\usepackage{hyperref}
\usepackage{amsmath}
\usepackage{amssymb}

\usepackage{cleveref}
\crefname{equation}{Eq.}{Eqs.}
\Crefname{equation}{Equation}{Equations}
\crefrangeformat{equation}{Eqs.~(#3#1#4--#5#2#6)}
\crefname{figure}{Fig.}{Figs.}
\Crefname{figure}{Figure}{Figures}
\crefname{table}{Table}{Tables}
\Crefname{table}{Table}{Tables}
\crefname{section}{Sect.}{Sects.}
\Crefname{section}{Section}{Sections}
\crefname{subsection}{Sect.}{Sects.}
\Crefname{subsection}{Section}{Sections}
\crefname{appendix}{Appendix}{Appendices}
\Crefname{appendix}{Appendix}{Appendices}

\usepackage{makecell}
\usepackage{tikz}
\usepackage{rotating}
\usetikzlibrary{decorations.pathmorphing}
\usetikzlibrary{arrows.meta}
\usetikzlibrary{bending}
\usetikzlibrary{external}
\usepackage{xcolor}
\definecolor{unamblue}{cmyk}{1 0.79 0.12 0.59}
\usepackage{color}
\usepackage{placeins}
\usepackage{multirow}

\hypersetup{
  colorlinks=true,%
  citecolor=unamblue,%
  filecolor=unamblue,%
	 linkcolor=unamblue,%
	 urlcolor=unamblue,
	 bookmarksnumbered=true,     
	 bookmarksopen=true,         
	 bookmarksopenlevel=1,       
	 pdfstartview=Fit,           
	 pdfpagemode=UseOutlines,
	 pdfpagelayout=TwoPageRight
	}

\numberwithin{equation}{section}

\def\Numer{\mathcal N}
\def\Denom{\mathcal D}

\begin{document}

\title{Finite Massless Pentaboxes}

\author{Gustavo Figueiredo}
\affiliation{Institut de Physique Théorique, CEA, CNRS, Université Paris--Saclay, F--91191 Gif-sur-Yvette cedex, France}
\affiliation{\textsf{\rm\sf gustavo.pereira-figueiredo@ipht.fr}}

\author{David A.~Kosower}
\affiliation{Institut de Physique Théorique, CEA, CNRS, Université Paris--Saclay, F--91191 Gif-sur-Yvette cedex, France}
\affiliation{\textsf{\rm\sf David.Kosower@ipht.fr}}

\author{Pavel P.~Novichkov}
\affiliation{Department of Physics and Astronomy, Ghent University, 9000 Ghent, Belgium}
\affiliation{\textsf{\rm\sf Pavel.Novichkov@UGent.be}}

\preprint{}

\begin{abstract}
We characterize the integrand numerators that 
give rise to locally finite or evanescent Feynman 
integrals for the massless pentabox. 
We provide compact expressions for the generators of the 
corresponding ideal in terms of an 
adapted momentum basis and also in terms of Gram determinants. 
We also compute the integrals corresponding to the lowest-rank
numerators in terms of polylogarithms 
using the \textsf{HyperInt\/} package, and in terms of pentagon functions.
\end{abstract}

\newcommand{\gramG}[2]{G\bigg(\begin{matrix}
        #1  \\
        #2
    \end{matrix}\bigg)}

\newcommand{\gram}[1]{G\big(\begin{matrix}
        #1
    \end{matrix}\big)}

\newcommand{\idealr}[2]{\langle \mspace{-10mu} \langle \mspace{3mu} #1 \mspace{+3mu} \rangle \mspace{-10mu} \rangle \mspace{+3mu} _{#2}}
\newcommand{\ideal}[1]{\langle \mspace{-10mu} \langle \mspace{3mu} #1 \mspace{+3mu} \rangle \mspace{-10mu} \rangle \mspace{+3mu}}

\newcommand{\maple}{\textsf{Maple\/} }
\newcommand{\eps}{\epsilon}
\newcommand{\pb}{pentabox}
\newcommand{\hyperint}{\textsf{HyperInt\/}{}}
\newcommand{\fire}{\textsf{FIRE\/} }
\newcommand{\PMI}{\textsf{PentagonMI\/}{}}
\newcommand{\PLT}{\textsf{PolyLogTools\/}{}}
\newcommand{\AMF}{\textsf{AMFlow\/} }

\newcommand{\define}{:=}
\newcommand{\collId}{Y}

\def\Ord{\mathcal{O}}
\def\Singular{\textsf{Singular\/}}

\newcommand{\GPL}{GPL}
\newcommand{\Sij}{S}
\newcommand{\ginac}{\textsf{GiNaC}}

\newcommand{\pavel}[1]{{\color{violet} #1}}
\def\draftnote#1{\textit{\color{orange} #1 --DAK}}

\tikzset{snake it/.style={decorate, decoration=snake}}
\tikzset{coil it/.style={decorate, decoration={coil, segment length = 4pt}}}
\tikzset{fermion/.style=-{Latex[length=6pt,bend,sep=1em,round]>[length=.5pt]}}
\tikzset{MassIns/.style=-{Rays[length=6pt,bend,sep=1em,round]>[length=0.5pt]}}
\tikzset{CounterTerm/.style=-{
Circle[white,length=5.5pt,bend,sep=-5pt]
Rays[length=5pt,bend,sep=-5.25pt,round]
Circle[open,length=6pt,bend,sep=.65em]>[length=0.5pt]}}
\tikzset{MassIns/.style=-{Rays[length=6pt,bend,sep=1em,round]>[length=0.5pt]}}
\tikzset{uzh/.style={thick,line cap =round}}

\maketitle
\flushbottom
\newpage

\input{Introduction}
\input{Setup}
\input{Organization}

\input{Results}

\input{IntResults}

\input{Conclusions}

\begin{acknowledgments}
We thank Erik Panzer, Ben Page and Leonardo de la Cruz for helpful discussions.
The work of DAK and GF was supported by the European Research Council (ERC) under the European Union’s research and innovation programme grant agreements ERC–AdG–
885414 (‘Ampl2Einstein’).
The work of PPN was supported by the
European Research Council (ERC) under the European Union’s Horizon Europe research and innovation program grant agreement 101078449 (ERC Starting Grant MultiScaleAmp).
Views and opinions expressed are however those of the authors only and do not necessarily reflect those of the European Union or the European Research
Council Executive Agency.
Neither the European Union nor the granting authority can be held responsible for them.
\end{acknowledgments}

\appendix

\input{DBExpr}
\input{IBPreduction}

\bibliographystyle{apsrev4-2mod}
\bibliography{refs}
\end{document}

%% file: Introduction.tex
\section{Introduction}
\label{ch:1}
\label{IntroductionSection}

Feynman integrals are key ingredients in calculations
of scattering amplitudes.  A reorganization of different
integrals along physically motivated lines may help simplify
the expression of amplitudes and possibly the calculation of
the integrals' rational coefficients as well.  
Important features
of scattering amplitudes that may benefit from an organization
that simplifies their presentation are infrared (IR) and
ultraviolet (UV) divergences. In dimensional
regularization both types of divergence appear as poles in the
regulator $\eps$. In reorganizing integrals according to their
IR divergences, we must first characterize integrals that
are either divergence-free, or integrals which inherently are
of $\Ord(\eps)$ and so can be ignored in the lowest-order 
observables in which they appear. 
In a recent article, Gambuti, Tancredi, and two of the 
authors (GKNT)~\cite{Gambuti:2023eqh} provided an algorithm for
systematically obtaining special integrands in Feynman
integrals.  These are special numerators which give rise to
either locally finite or locally evanescent ($\Ord(\eps)$) 
integrals.  These are integrals which have only integrable
singularities anywhere in loop-momentum space.  Integrals may
of course also be finite through cancellation of singularities
in different regions or through cancellation of overall 
evanescence with poles; we will not consider such integrals here.

The construction of finite integrals had been
considered previously in Refs.~\cite{vonManteuffel:2014qoa,%
vonManteuffel:2015gxa}, via subtraction in Refs.~\cite{Anastasiou:2022eym,Anastasiou:2018rib}, and more recently in 
Refs.~\cite{Salvatori:2024nva,delaCruz:2024xsm,%
Ma:2025mog,%
Dhani:2026cxx}.

In this article, we study the massless two-loop
five-point integral, \textit{aka\/} pentabox.
This integral provides a richer playground for the GKNT methods.
We apply them to obtain a complete classification of locally finite and evanescent
integrals (before any further simplifications
from integration-by-parts (IBP) relations). We refine the GKNT approach using adapted variables in the algebraic geometry computations.

Finite integrals allow computational methods not available for divergent integrals. We apply Panzer's 
\hyperint{}~\cite{Panzer:2014caa} to computing the lowest-rank finite and evanescent integrals. We cross-check the computation using a reduction to the pentagon functions of Chicherin, Gehrmann, Henn, and Sotnikov~\cite{Gehrmann:2018yef,%
Chicherin:2017dob,Gehrmann:2015bfy,Chicherin:2020oor}. We obtain two different representations which 
agree numerically, and also agree with \AMF \cite{Liu:2022chg} evaluations. 
These three integrals can serve as the top-level pentaboxes in an IBP basis for this topology.
\hyperint{} is a \maple implementation of the automated hyperlogarithmic integration algorithm of Brown~\cite{Brown:2008um,Brown:2009ta} and Panzer~\cite{Panzer:2015ida,Panzer:2014caa}. Giroux, Mizera, and Salvatori have recently presented \textsf{Subtropica}, a \textit{Mathematica} 
implementation~\cite{Giroux:2026tgd} of a similar algorithm with subtractions. 

In the next section, we
present our notation.  In \cref{ResultsOverview}, we
discuss the general features of the results.
In \cref{PentaboxNumerators}, 
we give explicit forms for the
numerators.  In \cref{Integrations}, we discuss the
integration procedure and features of the results;
the integration results are given in ancillary files.
We present our conclusions in \cref{Conclusions}.
 

%% file: Setup.tex
\section{Setup and Notation}
\label{ch:Setup}
\label{ch:2}

We largely follow the notations of 
Ref.~\cite{Gambuti:2023eqh}, writing two-loop 
dimensionally regulated Feynman integrals as,
\begin{equation}\label{eq:twoloop}
    I\left[\Numer(\ell_i)\right] = 
    \int d^D\ell_1 d^D\ell_2 \frac{\Numer(\ell_i)}
    {\Denom_1 \cdots \Denom_E} \,,
\end{equation}
where the $\Denom_e = q_e - m^2_e + i \varepsilon$
are the $E$ denominators of the integrals,
$\Numer(\ell_i)$ is a Lorentz-invariant numerator.  
We will leave the Feynman 
$i\varepsilon$ implicit. As usual, we take $D=4-2 \eps$. 

\begin{figure}[htb]
    \centering
    \includegraphics[width=0.35\linewidth]{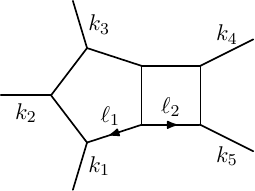}
    \caption{The massless \pb{} graph.}
    \label{fig:pb}
\end{figure}
\FloatBarrier

In this article, we study the massless five-point
integral, where all $m_e=0$.  
We show a diagrammatic representation of the integral
in \cref{fig:pb}.
Our task is to find all $\Numer(\ell_i)$ that
make this integral locally finite. 
By locally finite we mean integrals that could be 
performed strictly in $D=4$, that is, not involving any 
$\eps/\eps$ cancellations.
We take all external vectors to be strictly four-dimensional. We can reduce any external vector $Q_j$ to 
a linear combination of the external momenta, so that it 
suffices to consider numerators built out of the loop 
momenta $\ell_{1,2}$ and external momenta 
$k_{1\cdots5}$. The numerators are thus polynomials in the $\ell_i \cdot \ell_j$ and $\ell_i \cdot k_j$. We will denote by the term \textit{rank\/}
the degree in the $\ell_i$ of any monomial.
 As numerators are in general not homogeneous in rank,
we define the rank of a polynomial as the maximal
rank of any of its terms. 
We will call a numerator giving rise to a
locally finite integral a \textit{finite numerator\/}.
As a byproduct we will also obtain numerators which give rise to integrals of $\Ord(\eps)$, 
which we call \textit{evanescent numerators\/}.

The
propagator denominators are,
\begin{equation}
\label{eq:pbProps}
\begin{aligned}
    &\Denom_1 = \ell_1^2 \,,\quad
    \Denom_2 = (\ell_1 -k_1)^2 \,,\quad
    \Denom_3 =  (\ell_1 -K_{12})^2\,,\quad 
    \Denom_4 = (\ell_1 -K_{123})^2 \,,\\
    &\Denom_5 = (\ell_1 +\ell_2)^2 \,,\quad
    \Denom_6 = \ell_2^2 \,,\quad
    \Denom_7 = (\ell_2 -k_5)^2 \,,\quad
    \Denom_8 = (\ell_2 -K_{45})^2 \,.
\end{aligned}
\end{equation}
Here we have used the notation $K_{i \cdots j} = 
k_{i}+\cdots + k_{j}$.  The external momenta satisfy momentum conservation, are all
massless and taken to be outgoing. We use the usual Mandelstam invariants,
\begin{equation}
    K_{ij}^2 = s_{ij} \ .
\end{equation}
Not all the $s_{ij}$ are independent; we use the following independent set
\begin{equation}
     \Sij = \{s_{12},s_{23},s_{34},s_{45},s_{51}\} \,.
\end{equation}
For five-point kinematics these invariants remain independent in four dimensions. 

In using computational algebraic geometry tools
such as \Singular~\cite{Singular} in following sections,
we write the coefficients in the numerators in terms of $s_{12}$ and dimensionless parameters $\chi_{ij}$,
\begin{equation}\label{eq:chis}
    \chi_{ij} = \frac{s_{ij}}{ s_{12}}\,.
\end{equation}
This allows us to keep numerators homogeneous in the
set of variables 
$V_0 = \{\ell_i\cdot\ell_j,\ell_i\cdot k_j,s_{12}\}$. For explicit computations in \cref{PentaboxNumerators} we will use decompositions into collinear and orthogonal directions as variables. This simplifies the computations.

We will use Gram determinants,
\begin{align}
    G 
    \begin{pmatrix}
        p_1 & \cdots & p_R  \\
        q_1 & \cdots & q_R
    \end{pmatrix} & \define \mathrm{det}(2 p_i \cdot q_j), \\
    G\left(p_i \cdots p_R\right) &\define G 
    \begin{pmatrix}
        p_1 & \cdots & p_R  \\
        p_1 & \cdots & p_R
    \end{pmatrix}\ ,
\end{align}
where $\{q_i\}$ and $\{p_i\}$ are two sets of $R$ momenta in $D$ dimensions. When the momenta in the Gram matrix are the external momenta $k_i$, we will often replace the explicit momentum vectors by their index,
\begin{equation}
    \gramG{1 &  2 & 3}{\ell_1 & 2 & 3 } = G \begin{pmatrix}
        k_1 & k_2 & k_3  \\
        \ell_1 & k_2 & k_3
    \end{pmatrix} \ .
\end{equation}
We also introduce the following shorthand notation for the Gram determinant involving all four linearly independent external momenta:
\begin{equation}
    \Delta \define \gram{1 & 2 & 3 & 4} \ .
\end{equation}

%% file: Organization.tex
\section{Overview of the Finite Numerators}
\label{ResultsOverview}
\label{ch:3}
We use the methods introduced in Ref.~\cite{Gambuti:2023eqh} to obtain the vector space of finite numerators. 
The Landau analysis Ref.~\cite{Gambuti:2023eqh} yields all independent numerators
giving rise to finite integrals.  This vector space
has additional structure, as explained in Sect.~IV~D
of Ref.~\cite{Gambuti:2023eqh}.  Because multiplying 
any IR-finite numerator by a polynomial in the loop momenta yields another IR-finite numerator, the IR-finite
numerators naturally form an \textit{ideal\/}.  This
allows us to characterize the independent numerators
using generators of the ideal, and also to use
standard methods of algebraic geometry to analyze
general numerators\footnote{Imposing UV constraints  spoils the ideal structure, because the resulting limit on numerator rank means that the a high-rank numerator may not yield a finite integral in spite of belonging to the ideal of IR-finite numerators. We will work within the IR-finite ideal and impose UV constraints only at the end.}.
We will refer to the ideal of all IR-finite numerators of an integral $I$ as $\ideal{ I }$. We introduce
as well a notation for the (sub)ideal generated by IR-finite numerators
of rank up to $r$, which  we denote $\idealr{I}{r}$.

We seek to grade the ideal by rank.
We first form the Gr{\"o}bner basis of
lowest-rank numerators; for the pentabox ideal $\ideal{I_{\mathrm{pb}}}$,
these are of rank two. We perform
computational algebraic geometry calculations using 
\Singular. We take the set of 
$\ell_i \cdot \ell_j$, $\ell_i \cdot k_j$, and $s_{12}$ 
as variables, and use a weighted reverse degree 
lexicographic ordering where each variable is weighted 
by its rank.

\begin{table}[thb]
\renewcommand{\arraystretch}{1.1}
\setlength{\tabcolsep}{8pt}
\begin{tabular}{l|ccccccc}
Rank & 1 & 2 & 3 & 4 & 5 & 6 & 7 \\
\hline\hline
 IR-finite integrals & 0 & 3 & 36 & 211 & 850 & 2727 & 7496 \\
 IR- and UV-finite integrals & 0 & 3 & 36 & 211 & 782 & 2076 & 4235 \\
 Generators & 0 & 3 & 10 & 8 & 0 & 0 & 0 \\
 Top-level generators & 0 & 3 & 9 & 4 & 0 & 0 & 0\\
 \hline
 Evanescent integrals & 0 & 1 & 12 & 79 & 324 & 950 & 2109 \\
 Evanescent generators & 0 & 1 & 3 & 5 & 0 & 0 & 0 \\
 Top-level evanescent generators & 0 & 1 & 3 & 4 & 0 & 0 & 0
\end{tabular}

\caption{Counts of pentabox numerators and generators.
In the first two rows and the fifth row, the counts 
are inclusive, giving
the number of integrals up to and including
the given rank. In the remaining rows, the counts are of 
generators
newly arising at the given rank.}\label{CountsTable}
\end{table}
\FloatBarrier

The number of basis elements of lowest rank will necessarily be the same as the number of independent lowest-rank numerators. This initial Gr{\"o}bner basis will, however, also contain elements of higher rank. 
We then proceed to consider the remainders of the next rank numerators with respect to this Gr{\"o}bner basis, and then extract a linearly independent set. We do this by looking at the coefficients over a basis of all monomials present in the various polynomials, using finite-field values for the $\chi_{ij}$. 
For the pentabox, we find 33 remainders at rank three,
but only 10 are linearly independent. This means we need 10 additional generators in going from $\idealr{I_{\mathrm{pb}}}{2}$ to $\idealr{I_{\mathrm{pb}}}{3}$. At rank four, we find 172 remainders, but only 8 new generators. Beyond rank 4, we need no new generators. 
We summarize this analysis, along with the result of 
imposing the UV constraint, in the first three lines of 
\cref{CountsTable}.

It turns out that we can further distinguish different 
classes of generators. We can choose some generators to 
be proportional to a denominator. Such a generator will 
effectively correspond to an integral with a propagator 
omitted, that is a daughter integral.\footnote{It may 
not be a numerator for the daughter integral because not
all the momenta appearing in it are external momenta of 
the daughter.}
Such choices already appeared in the finite integrals 
presented in Ref.~\cite{Gambuti:2023eqh}, though the
authors there did not perform the analysis 
systematically. 

We can test such reducibility at rank $r$ by examining 
the linear independence of the remainders of the rank 
$r$ generators over the subideal $\idealr{I}{r-1}$ augmented in turn by each of the  denominators $\Denom_j$. A linear dependence signals the possibility of choosing one or more generators proportional to that denominator. The number of such factorizing generators is given by the dimension of the null space of the remainders.
We find one such reducible generator at rank three, 
and four at rank four. The remaining generators we call \textit{top-level} generators. We summarize these results in the 
fourth line of \cref{CountsTable}.

As explained in Ref.~\cite{Gambuti:2023eqh}, when a finite numerator vanishes upon taking a loop momentum to be strictly four-dimensional, the corresponding integral will be of $\Ord (\eps)$. One can find the subspace of evanescent numerators of the complete vector space of finite numerators, or of the set of generators, or of the set of top-level generators. We summarize these
results in lines five through seven of \cref{CountsTable}.

What about UV finiteness?  Weinberg's theorem \cite{Dyson:1949ha,Weinberg:1959nj} tells us that a Feynman integral is UV-finite if it, and all its subintegrations, have a negative superficial degree of divergence. By subintegrations, we mean all possible integrations of the loop momenta where at most $(L-1)$ loops of the original graph have been integrated over. An overall negative superficial degree of divergence restricts the highest possible numerator rank to be
\begin{equation}
    r = 2E-4L-1 \ .
\end{equation}
The \pb{} integral's superficial degree of divergence 
limits the numerator rank to $r_{max} = 7$.
Constraints arising from subintegrations do not change 
the maximal rank, but do further constrain the loop 
momentum monomials allowed in the numerator. We call
the constraints based solely on superficial degrees of
divergence \textit{strong\/} UV constraints.  They
start to exclude numerators at rank 5. 
The coefficients of some of the expected
UV divergences may turn out to vanish. Allowing
corresponding numerators would correspond to imposing a 
\textit{weak\/} UV constraint, which we will not 
consider in the present article.
The number of strongly UV- as well as IR-finite 
integrals is given on the second line of \cref{CountsTable}. 
The count of evanescent numerators given on the fifth line is after application of the strong UV constraint.
The counts of finite numerators include the corresponding
evanescent ones: the counts on the second line include 
those on the fifth line, those on the third line include
those on the sixth line, and those on the fourth line
include those on the seventh and last line.
Evanescent numerators which are UV divergent may yield
finite integrals through an $\eps/\eps$ cancellation.
Such integrals are not locally finite, but would be
interesting to investigate as they are linked to the
appearance of rational terms in scattering amplitudes.
Georgoudis and Page~\cite{Georgoudis:2026han} have 
studied them recently.

%% file: Results.tex
\section{Explicit Forms for Finite Generators}\label{PentaboxNumerators}
\label{sec:pbIdeal}

We find 38 different relevant pinch surfaces associated with the massless \pb{}.
The pinch surfaces form a nested structure.  Most of them are contained inside other surfaces,
the outermost ones being
the five ``single-collinear'' surfaces
\begin{equation}
  \label{eq:collinearLimits}
  \ell_1 \parallel k_1, \quad \ell_1 - k_1 \parallel k_2, \quad \ell_1 - K_{12} \parallel k_3, \quad \ell_2 \parallel k_5, \quad \ell_2 - k_5 \parallel k_4,
\end{equation}
as well as the ``single-soft'' surface \(\ell_1 + \ell_2 = 0\).
All other pinch surfaces are contained in one (or more) of these six.

In general, the IR power counting may be different for a parent surface and a subsurface, 
therefore subsurfaces need to be analyzed separately.
A direct check shows that, for the massless \pb{}, the single-soft surface is integrable in four dimensions, together with all its subsurfaces; the single-collinear surfaces and all their subsurfaces are logarithmically divergent.
This means that the numerators rendering our integral IR-finite are precisely those that vanish in the five collinear limits~\eqref{eq:collinearLimits}.

We can make the collinear limits manifest by splitting the loop momenta into their four- and \((-2 \epsilon)\)-dimensional parts as \(\ell_i = \bar{\ell}_i + \hat{\ell}_i\) and further parameterize the four-dimensional parts as follows:
\begin{equation}
  \begin{aligned}
    \bar{\ell}_1 &= x_1 k_1 + x_2 k_2 + x_3 k_3 + \ell_1^\perp, \\
    \bar{\ell}_2 &= x_4 k_4 + x_4 k_5 + \ell_2^\perp.
  \end{aligned}
\end{equation}
Here, \(\ell_1^\perp\) is the projection of \(\bar{\ell}_1\) onto the one-dimensional space orthogonal to \(k_1, k_2, k_3\); similarly, \(\ell_2^\perp\) is the projection of \(\bar{\ell}_2\) onto the two-dimensional space orthogonal to \(k_4, k_5\).

The \(x_i\) variables can be written in terms of Gram determinants as
\begin{equation}
  \label{eq:x-variables}
  \begin{gathered}
    x_1 = \frac{\gramG{\ell_1 \ 2 \ 3}{1\ 2\ 3}}{\gram{1 \ 2 \ 3}}, \quad
    x_2 = \frac{\gramG{1 \ \ell_1 \ 3}{1\ 2\ 3}}{\gram{1 \ 2 \ 3}}, \quad
    x_3 = \frac{\gramG{1 \ 2 \ \ell_1}{1\ 2\ 3}}{\gram{1 \ 2 \ 3}},
    \\
    x_4 = \frac{\gramG{\ell_2 \ 5}{4 \ 5}}{\gram{4 \ 5}}, \quad
    x_5 = \frac{\gramG{4 \ \ell_2}{4 \ 5}}{\gram{4 \ 5}}.
  \end{gathered}
\end{equation}
We can parametrize the \(\ell_j^\perp\) vectors 
explicitly in terms of scalars $\ell_{j,r}^\perp$, for example as follows,
\begin{equation}
  \label{eq:lperp-variables}
  \begin{aligned}
    \ell_{1,1}^\perp &=
    \ell_1^\perp \cdot k_4 = 
    \frac{\gramG{1 \ 2 \ 3 \ \ell_1}{1 \ 2 \ 3\ 4}}{\gram{1 \ 2 \ 3}} , \\[2mm]
    \ell_{2,1}^\perp &=
    \ell_2^\perp \cdot k_1 =
    \frac{\gramG{\ell_2 \ 4 \ 5}{1 \ 4 \ 5}}{\gram{4 \ 5}}, \quad
    \ell_{2,2}^\perp =
    \ell_2^\perp \cdot k_2 = 
    \frac{\gramG{\ell_2 \ 4 \ 5}{2 \ 4 \ 5}}{\gram{4 \ 5}},
  \end{aligned}
\end{equation}
and collect the $\hat\ell_i$ vectors into products,
\begin{equation}
\label{nuVariables}
  \nu_{ij} = \hat{\ell}_i \cdot \hat{\ell}_j = 
  \frac{\gramG{\ell_i \ 1 \ 2 \ 3 \ 4}
  {\ell_j \ 1 \ 2 \ 3 \ 4}}{\gram{1 \ 2 \ 3 \ 4}}\,.
\end{equation}

We now interpret \(V_1 = \bigl\{ x_i, \ell_j^\perp, \nu_{ij}\bigr\}\) as the complete set of independent variables for integrand numerators.
These variables are related to the scalar product variables \(V_0\) by a polynomial (more precisely, quadratic) map whose inverse is also polynomial.
Any numerator which is polynomial in scalar products is then also polynomial in the \(V_1\) variables, 
and vice versa.
More generally, the two sets of variables are equivalent from the algebraic geometry point of view.
We can thus describe the ideal of finite-integral 
numerators using the \(V_1\) variables, and translate it back to the scalar-product variables by mapping generators via~\cref{eq:x-variables,%
eq:lperp-variables,nuVariables}. 
(A similar quadratic change of variables was used to simplify algebraic geometry computations in the context of syzygy equations in Ref.~\cite{Coro:2025kha}.)

\begin{table}[thb]
\centering
\begin{tabular}{c | c}
Rank & Expressions\\
\toprule
 2 &\vbox to 14pt{}\vtop to 8pt{}
 $ \nu_{12}\,,\quad  \ell_{1,1}^\perp \ell_{2,1}^\perp\,,\quad  \ell_{1,1}^\perp \ell_{2,2}^\perp $\\
 \hline
 \multirow{3}{*}{\vbox to 14pt{}3} &\vbox to 14pt{}\vtop to 8pt{}
     $\nu_{11} \ell_{2,1}^\perp\,,\quad \nu_{11} \ell_{2,2}^\perp\,,\quad \nu_{22} \ell_{1,1}^\perp\,,$\\
   &\vbox to 8pt{}\vtop to 8pt{}
   $(x_1-1) x_2 \ell_{2,1}^\perp\,,\quad
   (x_1-1) x_3 \ell_{2,1}^\perp\,,\quad (x_2-1) x_3 \ell_{2,1}^\perp\,,\quad
   (x_5-1) x_4 \ell_{1,1}^\perp\,,$  \\ 
   &\vbox to 8pt{}\vtop to 8pt{}
   $(x_1-1) x_2 \ell_{2,2}^\perp\,,\quad
   (x_1-1) x_3 \ell_{2,2}^\perp,\ (x_2-1) x_3 \ell_{2,2}^\perp$  \\ 
   \hline
 \multirow{2}{*}{4} &\vbox to 14pt{}\vtop to 8pt{}
   \hspace{4mm}
   $\nu_{11}\nu_{22}\,,\quad \nu_{11} (x_5-1) x_4\,,\quad \nu_{22} (x_1-1) x_2\,,\quad 
   \nu_{22} (x_1-1) x_3\,,\quad 
   \nu_{22} (x_2-1) x_3\,, $\hspace{4mm}\\
   &$(x_1-1) x_2 (x_5-1) x_4\,,\quad 
   (x_1-1) x_3 (x_5-1) x_4\,,\quad 
   (x_2-1) x_3 (x_5-1) x_4$
\end{tabular}
\caption{Compact expressions for generators of the ideal of IR-finite \pb{} numerators.}\label{tab:pbGens}
\end{table}
\FloatBarrier

The advantage of the \(V_1\) variables is that ideals of numerators vanishing on individual collinear surfaces take a particularly simple form independent of the kinematic invariants.
The generators can be guessed easily:
\begin{equation}
  \begin{aligned}
    \collId_1 &= \left\langle x_2, x_3, \ell_{1,1}^\perp, \nu_{11}, \nu_{12} \right\rangle\,, \\
    \collId_2 &= \left\langle x_1 - 1, x_3, \ell_{1,1}^\perp, \nu_{11}, \nu_{12} \right\rangle\,, \\
    \collId_3 &= \left\langle x_1 - 1, x_2 - 1, \ell_{1,1}^\perp, \nu_{11}, \nu_{12} \right\rangle\,, \\
    \collId_4 &= \left\langle x_4, \ell_{2,1}^\perp, \ell_{2,2}^\perp, \nu_{22}, \nu_{12} \right\rangle\,, \\
    \collId_5 &= \left\langle x_5 - 1, \ell_{2,1}^\perp, \ell_{2,2}^\perp, \nu_{22}, \nu_{12} \right\rangle\,,
  \end{aligned}
\end{equation}
where \(\collId_i\) is the ideal of numerators canceling the collinear divergence associated with \(k_i\).
Then, the sought-after ideal of finite-integral numerators can be computed as 
the ideal intersection of the \(\collId_i\) using \Singular.
Again choosing a weighted monomial order consistent with loop-momentum degree, that is, assigning weight 2 to \(\nu_{ij}\) with the remaining variables taken to have weight 1, we obtain a Groebner basis that is naturally organized by rank.  We display it in \cref{tab:pbGens}.

Further intersecting the result with \(\left\langle \nu_{11}, \nu_{22}, \nu_{12} \right\rangle\) we find the ideal of evanescent numerators; its Groebner basis consists of the generators in \cref{tab:pbGens} proportional to \(\nu_{ij}\).

We have verified numerically that both the IR-finite and
the evanescent ideals constructed in this way coincide with the corresponding ideals obtained by fitting a linear ansatz using the method of Ref.~\cite{Gambuti:2023eqh}.

The set of generators in \cref{tab:pbGens} is not unique. We may use the above arguments to write an alternative set of generators, spanning the same ideal:

\begin{equation}
\label{eq:gramGens}
\begin{aligned}
  \mathrm{rank \ 2:}& \quad g_{pb,1}^{(2)}=\gramG{\ell_1 &  1 & 2 & 3}{4 &  1 & 2 & 3} \gramG{\ell_2 & 4 & 5}{1 & 2 & 3}\,, \quad   
  g_{pb,2}^{(2)}=\gramG{\ell_2 & 5 & 4 & 1}{\ell_1 &  1 & 2 & 3}\,,\\
  \mathrm{rank \ 2\ ev.:}& \quad g_{pb,3}^{(2)}=\gramG{\ell_1 &  1 & 2 & 3 & 4}{\ell_2 &  1 & 2 & 3 & 4}\,, \\
  \mathrm{rank \ 3:}
  & \quad  g^{(3)}_{pb} = \gramG{\ell_1 & 1 & 2 & 3}{4 & 1 & 2 & 3} \left( \ell_2 - k_5 \right)^2\,,
  \\
  & \quad \gramG{\ell_1 & 1 & 2}{\ell_1 & K_{12} & 3} \gramG{\ell_2 & 4 & 5}{1 & 4 & 5}\,,
  \quad \gramG{\ell_1 & 1 & 2}{\ell_1 & K_{12} & 3} \gramG{\ell_2 & 4 & 5}{2 & 4 & 5}\,,
  \\
  & \quad \gramG{\ell_1 & 1 & 2}{\ell_1 - k_1 & 2 & 3} \gramG{\ell_2 & 4 & 5}{1 & 4 & 5}\,,
  \quad \gramG{\ell_1 & 1 & 2}{\ell_1 - k_1 & 2 & 3} \gramG{\ell_2 & 4 & 5}{2 & 4 & 5}\,,
  \\
  & \quad \gramG{\ell_1 & 1 & 3}{\ell_1 - k_1 & 2 & 3} \gramG{\ell_2 & 4 & 5}{1 & 4 & 5}\,,
  \quad \gramG{\ell_1 & 1 & 3}{\ell_1 - k_1 & 2 & 3} \gramG{\ell_2 & 4 & 5}{2 & 4 & 5}\,,
  \\
  \mathrm{rank\ 3\ ev.{}:}
  & \quad \gramG{\ell_1 & 1 & 2 & 3}{4 & 1 & 2 & 3} \gram{\ell_2 & 1 & 2 & 3 & 4}\,,
  \\
  & \quad \gram{\ell_1 & 1 & 2 & 3 & 4} \gramG{\ell_2 & 4 & 5}{1 & 4 & 5}\,,
  \quad \gram{\ell_1 & 1 & 2 & 3 & 4} \gramG{\ell_2 & 4 & 5}{2 & 4 & 5}\,,
  \\
  \mathrm{rank \ 4:}
  &\quad \gramG{\ell_1 & 1 & 2}{\ell_1 & K_{12} & 3} \left( \ell_2 - k_5 \right)^2\,,
   \quad \gramG{\ell_1 & 1 & 2}{\ell_1 - k_1 & 2 & 3} \left( \ell_2 - k_5 \right)^2\,,
  \\
  &\quad \gramG{\ell_1 & 1 & 3}{\ell_1 - k_1 & 2 & 3} \left( \ell_2 - k_5 \right)^2\,,
\\  \mathrm{rank \ 4\ ev.:}
  & \quad \gram{\ell_1 & \ell_2 & 1 & 2 & 3 & 4}\,,
  \quad \gramG{\ell_1 & 1 & 2}{\ell_1 & K_{12} & 3} \gram{\ell_2 & 1 & 2 & 3 & 4}\,,
  \\
  & \quad \gramG{\ell_1 & 1 & 2}{\ell_1 - k_1 & 2 & 3} \gram{\ell_2 & 1 & 2 & 3 & 4}\,,
  \quad \gramG{\ell_1 & 1 & 3}{\ell_1 - k_1 & 2 & 3} \gram{\ell_2 & 1 & 2 & 3 & 4}\,,
  \\
  & \quad \gram{\ell_1 & 1 & 2 & 3 & 4} \left( \ell_2 - k_5 \right)^2\,.
\end{aligned}
\end{equation}
Here, the evanescent generators are marked with `ev.';
the other generators are not evanescent and 
are independent after modding out by the evanescent subideal.

\begin{figure}
    \centering
    \includegraphics[width=0.3\linewidth]{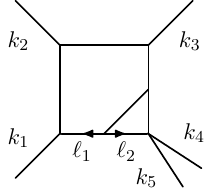}
    \caption{ The graph corresponding to the four-point one-mass Feynman integral one obtains when pinching the propagator $\mathcal D _7$ of the \pb{}. We refer to this graph as the \textit{massive beetle} graph.}
    \label{fig:beetle}
\end{figure}

The generators in \cref{eq:gramGens} proportional to $\Denom_7 = (\ell_2 - k_5)^2$ make manifest the reducibility of the generator set with respect to the propagator denominators. Generators reducible by $\mathcal D_7$ are related to the massive beetle integral, $I_{\textrm{mb}}$ depicted in \cref{fig:beetle}. 

The reducibility of $g_{pb}^{(3)}$ defined in \cref{eq:gramGens} suggests that the Gram 
determinant in terms of which it is written could be interpreted as a generator of $\idealr{I_{\mathrm{mb}}}{1}$. However, this numerator depends on $k_4$, which is not one of the external momenta of the graph. An independent analysis of $\idealr{I_{\mathrm{mb}}}{1}$ returns no rank-one finite numerators.  This discrepancy is partly
resolved by the observation that $I[g_{pb}^{(3)}]$
vanishes by Lorentz invariance, 
though integrals with polynomials
multiplying the generator need not vanish.

At rank four, all four top-level generators are evanescent. The remaining subideal is reducible by
$\Denom_7$, and an analysis of $\idealr{I_{\mathrm{mb}}}{2}$ reveals that the four reducible rank-four generators in \cref{eq:gramGens} can be taken to be
its generators. These results lead to the conclusion that in the limit $D\to 4$, $\idealr{I_{\mathrm{pb}}}{4}$ 
yields no nonvanishing top-level integrals.

%% file: IntResults.tex
\section{Integrating Finite Pentaboxes}
\label{Integrations}

Let us try computing the integrals corresponding to the lowest rank generators explicitly.
We use the forms given in \cref{eq:gramGens},
\begin{equation}
    I_i = I\bigl[ g^{(2)}_{pb,i} \bigr] \ .
\end{equation}

We use the \maple{}package \hyperint{}~\cite{Panzer:2014caa}, 
which could in principle be used for finite integrals generally;
and cross-check using integration-by-parts (IBP) reduction to the 
so-called \textit{pentagon functions\/} of
Refs.~\cite{Gehrmann:2018yef,Chicherin:2017dob,%
Gehrmann:2015bfy,Chicherin:2020oor}.  For the latter,
we use the implementation described in 
Ref.~\cite{Chicherin:2020oor}. 
In the limit $D\to 4$ we expect that $I_3 = 0$; this
emerges straightforwardly in the first approach,
but nontrivially in the second, where we compute it
as a cross check.

\subsection{Using \hyperint}\label{sub:hyperint}

\def\FeynNumer{\widetilde N}
\hyperint{} can integrate \textit{linearly reducible\/} 
integrals, that is whose parametric representation makes
possible the integration of a variable with
only linear dependence in the denominator 
at each step. 
Various authors~\cite{Bourjaily:2021lnz,%
Bourjaily:2019jrk,Bourjaily:2018aeq} have 
used \hyperint{}
to integrate multi-loop multi-leg Feynman integrals 
directly. To our knowledge the sorts of \pb{} integrals under consideration have not previously been computed 
this way.
To use \hyperint{}, we first express the integrands 
in terms
of Feynman parameters~\cite{Weinzierl:2022eaz,%
Heinrich:2008si,delaCruz:2024xsm,delaCruz:2025szs},
\begin{equation}\label{eq:param}
I[\Numer(\ell_i)] = 
   \Gamma\Bigl(E - \Bigl\lfloor \frac{r}{2} \Bigr\rfloor 
    -D\Bigr)\int d^E \alpha \; 
    \delta \Bigl( 1 - \sum _{e\in A}\alpha_e \Bigr) \,\mathcal U ^{\rho_u} \mathcal F^{\rho_f} \FeynNumer(\alpha_e)
\end{equation}
where the powers $\rho_u$ and $\rho_f$ are given by,
\begin{equation}
\begin{aligned}
    \rho_u &=E-3D/2-r\,, \\
    \rho_f &=D-E\,;
\end{aligned}
\end{equation}
$\FeynNumer(\alpha_e)$ is a polynomial in the $E$ 
Feynman parameters $\alpha_e$ with 
coefficients rational in the external invariants and
in $D$; $r$ is the highest rank among the terms in the polynomial $N(\ell_i)$ in \cref{eq:twoloop}.
The Cheng--Wu theorem allows us to choose the set $A$
to be any nonempty subset of $\{1 \dots E \}$.  We
choose $A$=$\{5\}$.

The polynomials
$\mathcal U$ and $\mathcal F$ are the first and second Symanzik polynomials, respectively.
Here, they are,
\begin{equation}
\begin{aligned}
    \mathcal U &= \alpha_{1} \alpha_{5} + \alpha_{2} \alpha_{5} 
    + \alpha_{3} \alpha_{5} + \alpha_{4} \alpha_{5} + \alpha_{1} \alpha_{6} 
    + \alpha_{2} \alpha_{6} + \alpha_{3} \alpha_{6} + \alpha_{4} \alpha_{6} 
    + \alpha_{5} \alpha_{6} \\
 & \quad + \alpha_{1} \alpha_{7} + \alpha_{2} \alpha_{7} + \alpha_{3} \alpha_{7} 
 + \alpha_{4} \alpha_{7} + \alpha_{5} \alpha_{7} + \alpha_{1} \alpha_{8} 
 + \alpha_{2} \alpha_{8} + \alpha_{3} \alpha_{8} + \alpha_{4} \alpha_{8} 
 + \alpha_{5} \alpha_{8}\,,\\
    -\mathcal F &= \alpha_{3} (\alpha_{5} \alpha_{6} + \alpha_{1} (\alpha_{5} 
    + \alpha_{6} + \alpha_{7} + \alpha_{8})) s_{12} 
    + \alpha_{2} \alpha_{5} \alpha_{7} s_{51} \\
 & \quad +\alpha_{2} (\alpha_{5} \alpha_{8} + \alpha_{4} (\alpha_{5} + \alpha_{6} 
 + \alpha_{7} + \alpha_{8})) s_{23} 
 + \alpha_{3} \alpha_{5} \alpha_{7} s_{34}  \\
 & \quad + \big(\alpha_{6} ((\alpha_{2} + \alpha_{3} + \alpha_{5}) \alpha_{8} 
 + \alpha_{4} (\alpha_{5} + \alpha_{8}))\\
 & \qquad \quad + 
    \alpha_{1} ((\alpha_{5} + \alpha_{6}) \alpha_{8} + \alpha_{4} (\alpha_{5} 
    + \alpha_{6} + \alpha_{7} + \alpha_{8}))\big) s_{45}\,.
\end{aligned}
\end{equation}

\Cref{eq:param} allows us to express the $I_i$ 
in parametric representation. For rank-two \pb{} 
integrals in $D=4$, we have $\rho_u=0$ and $\rho_f=-4$,
\begin{equation}
\begin{aligned}\label{eq:paramIntDefs} I_i = 
I\bigl[g^{(2)}_{pb,i}\bigr] = 
2\Delta\int d^8\alpha\;\delta\bigl(1-\alpha_5 \bigr) \,
\mathcal F^{-4}\FeynNumer\bigl[g^{(2)}_{pb,i}\bigr] \,,
\end{aligned}
\end{equation}
where we have pulled out a common factor of $\Delta$.
The Feynman-parameter numerators are,
\begin{align}
    \FeynNumer\bigl[g^{(2)}_{pb,1}\bigr]&=3\,
    \bigl( \alpha_{5}^2 \alpha_{7}\alpha_{3}\,c_{11}
    + \alpha_{5}^2\alpha_{7}\alpha_{2} \,c_{12}\bigr)\, ,
    \label{eq:n1}\\    \FeynNumer\bigl[g^{(2)}_{pb,2}\bigr]&=
    \frac{c_{21}}{\Delta} \alpha_{5}\mathcal F +3  \alpha_{5}^2 \alpha_{7} \alpha_{3} \,,
    \label{eq:n2}\\
    \FeynNumer\bigl[g^{(2)}_{pb,3}\bigr]&=0
    \label{eq:n3}\,.
\end{align}
The coefficients $c_{ij}$ appearing here
are, 
\begin{equation}
\begin{aligned}
    c_{11} &= \big((s_{23}-s_{51}) s_{12}^2+s_{45} (s_{45} (s_{34}-s_{51})-s_{23} s_{34})\\
     &\qquad-s_{12} (s_{23} (s_{34}+s_{45})+s_{45} (s_{34}-2 s_{51}))\big)\,,\\
    c_{22} &= ((s_{23}-s_{45}) (s_{34} (s_{23}-s_{45})+s_{45} s_{51})\\
    &\qquad+s_{12} (s_{23} (s_{45}-s_{23})+(s_{23}+s_{45}) s_{51}))\,,\\
    c_{21} &= (s_{12} (s_{23} (s_{45}-s_{23})+(s_{23}+s_{45}) s_{51})\\& \qquad+(s_{23}-s_{45}) (s_{23} s_{34}+s_{45} (s_{51}-s_{34})))\,.
\end{aligned}
\end{equation}

The numerator for $I_3$ is proportional to $\eps$, 
and hence vanishes in four dimensions.  This integral
will not vanish as trivially when reducing to 
pentabox functions, as we will see in the next
subsection.  Were we instead to compute it in 
$D=4-2\eps{}$, we would obtain,
\begin{equation}
    I_3^{(D)} =-2\eps{}\Gamma(3+2\eps{}) \Delta
    \int d^8 \alpha \; 
    \delta \bigl( 1 - \alpha_5 \bigr) 
    \mathcal U ^{3\eps{}} 
    \mathcal F^{-2(2+\eps{})}
    \,\alpha_5 \mathcal F\, .
\end{equation}
The limit of $I_3^{(D)}/\eps$ does give rise to
a potentially nonzero finite integral 
as $\eps\rightarrow 0$,
\begin{equation}
    \hat I^{(4)}_{3}\define
    \lim_{\eps{}\to 0 } 
    \Bigl(\frac{1}{\eps{}}I_3^{(D)}\Bigr) 
    = -4\Delta\int d^8 \alpha \;
    \delta \bigl( 1 - \alpha_5 \bigr) 
    \mathcal F^{-3}\, \alpha_5\,.
\end{equation}
We expect this integral to be of transcendental weight
five.  Its integrand
also appears as the term 
proportional to $\mathcal F$ in \cref{eq:n2}. 
As we will see, the integrals $I_1$ and $I_2$ are of
uniform transcendental weight four.  The presence
of these terms in the parametric representation of $I_2$
implies that all weight-five functions arising from
integrating the terms in \cref{eq:n2} must ultimately
cancel. We have checked that this is indeed the case. 
A similar cancellation occurs between the 
terms of $I_1$ in \cref{eq:n1}.

It turns out that naively the integrands are 
\textit{not\/} linearly reducible; they only
become linearly reducible if we take one of the basic
constants to be $\sqrt{\Delta}$ rather than $\Delta$.
We can proceed in two ways: work in an algebraic
field extension by $\sqrt{\Delta}$ (equivalently, 
factor the offending denominator by hand into two
factors, each dependent on $\sqrt{\Delta}$); or
change to variables which rationalize all expressions
and so make the integrand strictly linearly reducible.
The momentum twistor variables of 
Ref.~\cite{Badger:2013gxa} accomplish this goal.
They fully rationalize the integral's 
alphabet\footnote{We thank Eric Panzer for pointing this out, as well as for other helpful comments and code snippets which allowed us to utilize \hyperint{}
effectively.}. We convert back to the $s_{ij}$ 
after \hyperint{} completes. 
Ref.~\cite{Bourjaily:2018aeq} used 
momentum-twistor variables to rationalize square roots arising in the direct integration of several multi-loop hexagon, heptagon, and octagon integrals. Those integrals sometimes received contributions from certain finite \pb{} daughter diagrams as subintegrations, which the authors analyzed systematically. However, they
did not consider the massless \pb{}. 

In the ancillary files \texttt{data/int1HI.m}, and \texttt{data/int2HI.m} we present the results for the integrals $I_{1}$ and $I_{2}$ respectively in
terms of Goncharov polylogarithms (\GPL s) in the $s_{ij}$.
The result is valid in the Euclidean region,
\begin{equation}
    s_{ij} > 0.
\end{equation}
This region of validity is not further constrained by the surface defined by $\Delta = 0$;
we have checked this numerically.
The \GPL{}s on which the expressions depend can be evaluated numerically using \PLT{}~\cite{Duhr:2019tlz} or other codes. We did not use
the \textsf{fibrationBasis\/} routine in \hyperint{}.
As we will discuss further in 
\cref{sec:validation}, this omission
along with structural features of the result mean
that further simplification is possible, which we
defer to future work.

As a warm-up for the pentabox computation, we also used
\hyperint{} to integrate a rank-two finite double-box 
considered in Ref.~\cite{Gambuti:2023eqh}. That result
is given in Appendix~\ref{app:db}.

\subsection{Evaluation in Terms of Pentagon Functions}
In this subsection, we describe the second approach
to integrating the finite \pb{} integrals,
expressing them in terms of weight-four functions $F_\ell$:
\begin{equation}\label{eq:pf1}
    I_i = \sum_m f_m(\Sij) F_m\,.
\end{equation}
The $f_m(\Sij)$ are algebraic functions of the kinematic variables, and $F_m$ are built out
of the pentagon functions defined in Ref.~\cite{Chicherin:2020oor}. 

The authors of Ref.~\cite{Chicherin:2020oor} consider the differential equations and canonical bases for all planar and non-planar one- and two-loop massless five-point Feynman integrals, including all permutations of the external legs for each diagram.
The solutions to said differential equations are then given in terms of a basis of linearly independent irreducible iterated integrals over the corresponding alphabet. This alphabet contains the \pb{} alphabet as a subset. By `irreducible', the authors mean special functions at transcendental weight $w$ that cannot be stated as a product of functions with transcendental weight $w' < w$. This basis is given the name of pentagon functions. Because the pentagon functions were obtained as the solution of differential equations, they satisfy expected analyticity properties of Feynman integrals such as first-entry conditions, Steinmann relations, and integrability by 
construction~\cite{Hannesdottir:2024hke,%
Caron-Huot:2020bkp,Hannesdottir:2021kpd}.
By considering all permutations of the external legs, the authors define a map which relates the solution of the integrals in a fixed physical scattering region (chosen to be the $s_{12}$ scattering channel characterized by $12\to345$ scattering) to any other physical scattering region. In the present article, we will focus on the $s_{12}$ scattering channel 
in order to obtain finite analytic solutions. We will also make use of the ability to evaluate the master integrals in any of the physical scattering regions in order to validate analytic continuations of the results discussed in \cref{sub:hyperint}.

To obtain \cref{eq:pf1}, we first perform an IBP reduction of the finite integrals $I_i$ into the canonical master integrals of the \pb{} family as chosen in in Ref.~\cite{Chicherin:2020oor}. We denote the set of 61 canonical master integrals by $J_i$, and perform the reduction using the 
\fire software package~\cite{Smirnov:2025prc}. This provides expressions for our finite integrals as a linear combination of canonical masters, 
\begin{equation}\label{eq:ItoCan1}
        I_i = \sum_{j=1}^{61}C_{ij}J_j \ .
\end{equation}
We relegate a detailed description of the IBP reduction procedure to Appendix~\ref{ap:ibp}. Through this direct computation, we learn that the matrix $C_{ij}$ has the maximum possible rank of $3$, and as such the finite integrals under consideration here are linearly independent under IBP identities.

The finite integrals as expressed in \cref{eq:ItoCan} can be evaluated numerically 
in any physical scattering region with the packages provided in~\cite{Chicherin:2020oor}. 
However, the integrals $J_i$ are not manifestly finite in the limit $D\to4$. Replacing the $J_i$ by their expressions in terms of pentagon functions we obtain \cref{eq:pf1}. In that form, the integrals $I_i$ are linear combinations of products of pentagon functions up to transcendental weight 4.  The result is free of $\eps{}$, and manifestly finite in four dimensions.

The analytic definitions of the pentagon functions and the \PMI{} software package required to numerically evaluate the finite integrals are given in Ref.~\cite{Chicherin:2020oor}. 
With \PMI, we can evaluate the finite integrals in the kinematic region associated with the $12 \to 345$ scattering channel defined by the following constraints on the kinematics,
\begin{equation}
\begin{aligned}
    s_{12},\ s_{34},\ s_{45} > &\  0 \ ,\\
    s_{51},\ s_{23}, \ <& \ 0 \ ,\\
    \Delta <&\  0 \ , \\ \mathrm{Im}\left\{\sqrt{\Delta}\right\} <& \  0 \,.
\end{aligned}
\end{equation}
In the ancillary files \texttt{data/int1PF.m} and \texttt{data/int2PF.m} we present the results for the integrals $I_{1}$ and $I_{2}$ respectively in terms of pentagon functions.

\subsection{Validation and Analytic Structure}\label{sec:validation}

The independence of \cref{eq:pf1} with respect to $\eps{}$ is already a nontrivial check of our computation, as the cancellation of terms singular
in $\eps$ in the pentagon function expressions happens analytically.

To further validate our results, we perform several numerical checks. With the \hyperint{} result, 
it turns out that one can analytically continue away
from the Euclidean region and cross threshold singularities simply by giving the Mandelstam invariants a small imaginary part. We have used this freedom to check the agreement between the \hyperint{} result and the expression in terms of pentagon functions in the $12\to 345$ scattering channel. We can further compare analytic continuations of the \hyperint{} result with evaluations of the finite integrals in different physical scattering channels using \cref{eq:ItoCan1} and \PMI.
We have explicitly checked that the choice
\begin{equation}
    s_{ij}\to s_{ij}+i\varepsilon \ ,
\end{equation}
leads to numerical agreement between the two expressions 
for the finite integrals in the five $2\to 3$ scattering 
regions with center-of-mass energy $s_{ij} \in S$. We 
perform further numerical checks with evaluations using 
\AMF{}\cite{Liu:2022chg}, and find agreement. 
\Cref{tab:numEvals} displays a summary of these 
numerical evaluations, and the methods used for 
each check was achieved.

\begin{table}[hb]
\begin{center}
\begin{tabular}{c|c|c|c|c|c|c}
      &Euclidean &$12 \to 345$ & $23 \to 145$
     & $34 \to 125$
     & $45 \to 123$
     & $51 \to 234$ \\
     \hline
     \PMI{} &  & $\checkmark$ & $\checkmark$ & $\checkmark$ & $\checkmark$ & $\checkmark$ \\ \hline
     \hyperint{} & $\checkmark$ & $\checkmark$ & $\checkmark$ & $\checkmark$ & $\checkmark$ & $\checkmark$ \\ \hline \AMF{} & $\checkmark$ & $\checkmark$ & & & 
\end{tabular}
\caption{A summary of numerical evaluations of the finite integrals computed, in which kinematic regions, and by which methods.}\label{tab:numEvals}
\end{center}
\end{table}

In the following discussion we rely on the definition of the \pb{} alphabet given in Ref.~\cite{Gehrmann:2015bfy}, but remark that this choice is in conflict with the alphabet definition of Ref.~\cite{Chicherin:2020oor}. This is because in the former reference, only the planar alphabet is used, where as in the latter, letters which are exclusive to nonplanar graphs are also included. This results in an ambiguity: letters $W^{p}_{20}-W^{p}_{26}$ of the planar alphabet map to letters $W^{np}_{26}-W^{np}_{31}$ of the nonplanar alphabet. However, this ambiguity is only superficial as it rarely explicitly arises in our analytic expressions: the only case being $W^{p}_{26}=W^{np}_{31}=\Delta$. Since we work exclusively with the planar graph we use $W_{26}=\Delta$, but emphasize in particular that this choice does not reflect the choice made in the definitions of the pentagon functions in \PMI.

We now remark on the analytic properties of our results. The results achieved with \hyperint{} contain \GPL s with arguments that do not appear to factorize in the \pb{} alphabet. The nonfactorizing
expressions consist of  arguments which involve the single anomalous polynomial factor
\begin{equation}
    \tilde w = s_{34} s_{45}+(s_{12}-s_{45}) s_{51}\ .
\end{equation}
This has been observed 
previously in integrations of Feynman integrals ~\cite{Bourjaily:2021lnz,Bourjaily:2019jrk}. 
Ref.~\cite{Bourjaily:2021lnz} suggests the possibility that $\tilde w$ may be eliminated by expressing the integration result in terms of a fibration basis. We observed that doing so in terms of twistor variables with the tools available in \hyperint{} complicates the transformation back into Mandelstam variables, where the Euclidean region is most clearly manifest. We therefore chose to continue working in terms of a set of Goncharov polylogaritms which are over-complete. We note that going into a fibration basis did not otherwise provide a significantly simpler result by the metrics of~\cref{tab:summaryHI}.
Furthermore, these polylogarithms do not manifestly satisfy other analytic consistency conditions such as the first entry conditions (even modulo beyond-the-symbol terms). The following alphabet letters appear in the first entries of the \GPL s appearing in the expressions for both of the finite integrals:
\begin{equation}\label{eq:badLetters}\mathcal W=
        \left\{W_{1}, W_{2}, W_{3}, W_{4}, W_{5}, W_{11}, W_{12}, W_{13}, W_{14}, W_{16}, \
W_{17}, W_{20}, W_{21}, W_{24}, W_{25}\right\} \ .
\end{equation}
The first entry conditions dictate that only the $W_i$ for $\left\{ 1, \dots,5\right\}$ should 
appear in the first entry. 

We would like to validate these analyticity conditions from within the Euclidean region. We do so by probing for the presence of the expected analytic behavior numerically. 
All numerical evaluations are performed with the \hyperint{} result by evaluating the \GPL s with the interface to \ginac~\cite{Bauer:2000cp} implemented in \PLT. This allows us to evaluate our integrals numerically at points within the Euclidean region that incrementally approach the zeros of the alphabet letters in \cref{eq:badLetters}. 
More specifically, we fix a random phase space point $S_0 = \left\{s_{12}^{(0)},\dots,s_{51}^{(0)}\right\}$ in the Euclidean region satisfying 
\begin{equation}
    W_i(\Sij_0) = 0\ , 
\end{equation}
for some $W_i \in \mathcal W$. We then approach this point by defining a sequence of points $\Sij_n$ such that we vary only one of the the kinematic variables $s_{\bullet}$ in $\Sij$ on which the given $W_i$ depends. This $\Sij_n$ is defined such that it approaches $\Sij_0$ as $n$ increases, by \begin{equation}
S_n = \left\{s_{12}^{(0)},\dots,s^{(0)}_{\bullet}\left(1+\frac{1}{10^{n/2}}\right),\dots,s_{51}^{(0)}\right\} \ .
\end{equation}
In practice, we restrict the point such that $-s^{(0)}_{ij} \sim \Ord(10)$ and limited ourselves to a maximum $n$ value of $50$.

We find agreement with the expected analytic properties: the expressions are indeed finite in kinematic limits where $W_i(\Sij_0)=0$ for $W_i\in \mathcal W$ and $i\ne \{1,\dots,5\}$. Similarly,
for limits where $W_i(\Sij_0)=0$ and $i = \{1,\dots,5\}$, we find that the integrals diverge as a power of a logarithm. We did not determine the power numerically.

\begin{table}[ht]
\begin{center}
\renewcommand{\arraystretch}{1.1}
\setlength{\tabcolsep}{8pt}
\begin{tabular}{c|c|c|c|c}
      Integral & terms & \makecell{distinct\\ functions} &  \makecell{distinct\\ function\\ arguments} & \makecell{independent\\ coefficients} \\ \hline\hline
     $I_1$ & 1126 & 1207 & 33 & 5\\
     $I_2$ & 1099 & 1191 & 32 & 5\\
\end{tabular}
\caption{Summary of analytic properties of the integrated results for integrals $I_1$ and $I_2$ achieved with \hyperint{}. The relevant functions are Goncharov polylogarithms, not including transcendental constants. The number of terms is defined as the number of distinct monomials in the special functions and transcendental constants. The number of independent coefficients refers to the number of linearly independent algebraic functions which appear in the coefficients of the monomials.}\label{tab:summaryHI}
\end{center}
\end{table}

The result in terms of the pentagon functions also does not manifestly satisfy the first entry conditions. The choice of basis functions that contain logarithms of letters allowed in the nonplanar case but disallowed in the planar case may be responsible.

\begin{table}[ht]
\begin{center}
\renewcommand{\arraystretch}{1.1}
\setlength{\tabcolsep}{8pt}
\begin{tabular}{c|c|c|c}
      Integral & terms & \makecell{distinct\\ special\\ functions}  & \makecell{independent\\ coefficients}\\ \hline\hline
     $I_1$ & 2665 & 301 & 6\\
     $I_2$ & 2665 & 301  & 5\\
     \hline
     $J_{59}$ & 359 & 288 &\\
     $J_{60}$ & 72 & 61  & \\
     $J_{61}$ & 72 & 61 &\\
\end{tabular}
\caption{Summary of analytic properties of the integrated results for integrals $I_1$ and $I_2$ in terms of pentagon functions, the relevant special functions. The number of terms is defined as the number of distinct monomials in the special functions and transcendental constants. The number of independent coefficients refers to the number of linearly independent algebraic functions which appear in the coefficients of the monomials.}\label{tab:summaryPF}
\end{center}
\end{table}
\FloatBarrier

In \cref{tab:summaryHI,tab:summaryPF} we list the number of terms appearing in each integral, defined to be the number of independent monomials in the set of transcendental functions and constants. While both sets of expressions have thousands of terms, we note that the kinematic coefficients of each term exhibit a simple structure: there are only a maximum of 6 linearly independent algebraic functions in terms of which all coefficients in each expression can be stated. The respective coefficient counts for each integral in each of the independent computations are detailed in the respective tables.

The two features discussed above (nonmanifestness of first-entry conditions and appearance of nonletters)
as well as the appearance of large integer coefficients in the pentagon-function results suggest that a 
simpler representation than either result should exist.
For example, reexpressing our result in terms of a tailored basis of logarithmic special functions which minimally reproduce the functions' symbol and imposing the above analyticity constraints should yield
a representation in which spurious singularities cancel.
The complicated forms may also be a consequence of our choice of ideal generator. We leave an investigation of further simplification of our results to future work.

During our explorations we noticed that the linear combination of canonical master integrals $J_i$ given by
\begin{equation}\label{eq:simplefinite}
    \frac{3}{2} J_{60}- J_{61}
\end{equation}
is finite when written in terms of pentagon functions, and involves only three transcendental weight-four
special functions. While this integral is not guaranteed to be \textit{locally} finite, it would be interesting to see if it admits a simple representation in terms of the numerators of \cref{eq:gramGens}.

We can search for such linear combinations 
systematically: we take arbitrary linear combinations 
of canonical master integrals with rational number 
coefficients,
\begin{equation}
    \sum_{i=1}^{61} a_i J_i \ ,
\end{equation}
and demand that the $\eps^{-m}$ terms in the
coefficients of the pentagon functions vanish ($m>0$).
This yields a set of linear constraints on the numerical coefficients $a_i$. 
 By construction, any integral written as a linear 
combination satisfying said constraints will be a pure, uniform transcendentality, finite integral. Solving for these constraints yields only one constraint relating top-level \pb{} integrals (that is, integrals involving the $J_i$ for $i=\{59,60,61\}$),
 \begin{equation}
     \frac{a_{22}}{3} - \frac{a_{26}}{3} - \frac{a_{43}}{3} - \frac{a_{48}}{3} - \frac{a_{55}}{3} -  \frac{2a_{60}}{3} -a_{61}=0.
 \end{equation}
 This constraint is satisfied by the linear combination of~\cref{eq:simplefinite}, as well as the by IBP identities relating  $I_3$ to the canonical basis. Integrals $I_1$ and $I_2$ are not pure integrals and we do not expect them to be captured by such relations.

%% file: Conclusions.tex
\section{Conclusions}
\label{Conclusions}

In this article, we have applied the techniques of
GKNT~\cite{Gambuti:2023eqh} to the pentabox,
a five-point two-loop integral.  We obtained
explicit expressions for the generators of
numerators yielding finite and evanescent
integrals.  As was the case
for the integrals studied
in Ref.~\cite{Gambuti:2023eqh}, the number of generators 
is vastly smaller than the number of independent
numerators. As is true for the planar and nonplanar double-boxes, all generators are of rank four
or lower.  We also separated generators which contain
a denominator factor, so that the corresponding
integral is a daughter integral. We used a set of  collinear and orthogonal variables to simplify the computation leading to explicit expressions for the generators. The study of the pentabox suggests that the technology can be applied to yet more complicated integrals without undue difficulty. It will be interesting to explore whether five-point amplitudes can be simplified using the finite basis described here and the singularity-free approach of Ref.~\cite{DeAngelis:2025agn}.

We evaluated integrals with the three lowest-rank
finite numerators (of which one is evanescent) using Panzer's
\hyperint \cite{Panzer:2014caa}. We cross-checked this evaluation via reduction
to pentagon functions~\cite{Chicherin:2020oor,Chicherin:2017dob,Gehrmann:2018yef}. Both approaches yield lengthy
but consistent polylogarithmic expressions.  As 
explained in \cref{sec:validation}, it is likely
that both expressions could be simplified further.

%% file: DBExpr.tex
\section{A Finite Double Box}\label{app:db}

\def\db{\textrm{db}}
\def\Li{\mathop{\textrm{Li}}\nolimits}
We used \hyperint{} to compute the finite double box integral considered in Ref.~\cite{Gambuti:2023eqh}, corresponding to the first rank 2 integral in Eq.~(A.2). We refer to Ref.~\cite{Gambuti:2023eqh} for the kinematic conventions and propagator definitions.

Its parametric representation is given by 
\begin{equation}\label{eq:DBparam}
I_{\db} =
 \int d^7 \alpha \; 
    \delta \Bigl( 1 - \sum _{e\in A}\alpha_e \Bigr) \,\mathcal U ^{-1} \mathcal F^{-3} \FeynNumer_{DB}(\alpha_e) \ ,
\end{equation}
with Symanzik polynomials
\begin{equation}
\begin{aligned}
     \mathcal U & = \alpha_{1} \alpha_{4} + \alpha_{2} \alpha_{4} + \alpha_{3} \alpha_{4} + \alpha_{1} \alpha_{5} + \alpha_{2} \alpha_{5} + 
 \alpha_{3} \alpha_{5} + \\ &\alpha_{4} \alpha_{5} + \alpha_{1} \alpha_{6} + \alpha_{2} \alpha_{6} + \alpha_{3} \alpha_{6} + 
 \alpha_{4} \alpha_{6} + \alpha_{1} \alpha_{7} + \alpha_{2} \alpha_{7} + \alpha_{3} \alpha_{7} + \alpha_{4} \alpha_{7} \ , \\
 \mathcal F & =-s\Big((\alpha_{5}(\alpha_{3}\alpha_{4} + 
      (\alpha_{2} + \alpha_{3} + \alpha_{4})\alpha_{7}))+ \\& \qquad 
   \alpha_{1}((\alpha_{4} + \alpha_{5})\alpha_{7} + 
     \alpha_{3}(\alpha_{4} + \alpha_{5} + \alpha_{6} + 
       \alpha_{7}))\Big) - 
 t \ \alpha_{2}\alpha_{4}\alpha_{6}\ ,
\end{aligned}
\end{equation}
and
\begin{equation}
    \FeynNumer_{\db} = -4 \alpha_{2} \alpha_{4}^2 \alpha_{6} s t (s + t) - 
 2 \mathcal F \alpha_{4} s (s + 2 t)\ .
\end{equation}
We choose $A = \{3\}$ by the Cheng--Wu theorem. After integrating we use the \textsf{fibrationBasis\/} routine to express the result in terms of a basis of special functions (see Refs.~\cite{Panzer:2014caa,Brown:2008um} for details) with the assumption $t<s$ in accordance with the Euclidean region. The result, in terms of multiple polylogarithms is
\begin{equation}
\begin{aligned}
  I_{\db}= & -\frac{11}{90} \pi^4 + \frac{2}{3}\pi^2 \Li_{2}\left(-x\right) + 
  \frac{2}{3} \pi^2 \Li_{(1, 1)}\left(1, -x\right) + \\ & 
  4 \Li_{(1, 3)}\left(1, -x\right) + 
  4 \Li_{(1, 2, 1)} \left(1, 1, -x\right) + 12\zeta_3 - 
  12 \Li_{1}\left(-x\right) \zeta_3 \ ,
  \end{aligned}
\end{equation}
where $x = t/s$.

%% file: IBPreduction.tex
\section{Reduction to Canonical Master Integrals}\label{ap:ibp}

In this appendix we describe in detail the IBP reduction procedure to express the finite integrals in terms of canonical master integrals. Our ultimate goal is to produce \cref{eq:pf1}. We begin by considering the full massless \pb{} integral family. It defines a 61-dimensional vector space of integrals, but we will make use of several distinct (and sometimes overcomplete) sets of integrals in our computation. Specifically, the canonical basis as given in~\cite{Chicherin:2020oor} consists of 61 basis integrals expressed in terms of 70 intermediate integrals. \fire will choose at most 61 of these 70 integrals as master integrals in the process of IBP reduction. In addition, we aim to perform all reductions numerically, and target a minimal set of functions for rational function reconstruction. As such, it will be beneficial to normalize away any square roots in the canonical basis definition. 

Let us explicitly define these bases and the linear transformations among them.
We have,
\begin{itemize}
    \item $I_i$: the set of finite integrals we wish to express in terms of canonical master integrals,
    \item $J_j$: the set of canonical integrals,
    \item $J_j^{(aux)}$: the set of integrals obtained by removing all algebraic normalizations in the definition of $J_j$,
    \item $B_l$: the set of intermediate integrals in terms of which $J_j$ are given,
    \item $L_j$: the set of master integrals \fire will select in the IBP reduction procedure.
\end{itemize}
Expressing the finite integrals in the canonical basis can be stated as
    \begin{equation}\label{eq:ItoCan}
        I_i = \sum_{j=1}^{61}C_{ij}J_j \ ,
    \end{equation}
where $C$ is an algebraic $3 \times 61$ matrix of IBP identities. Likewise, we have
\begin{equation}
    J_j = \sum_{j=1}^{61}R_{jk}J_k^{(aux)} \ ,
\end{equation}
where $R$ is a $61 \times 61$ diagonal matrix consisting of the algebraic part of the kinematic normalizations present in the definition of $J_j$. We prefer to work with $J_i^{(aux)}$ to avoid performing functional reconstruction of expressions involving algebraic elements. Similarly, $B_l$ is defined by
\begin{equation}
    J^{(aux)}_k = \sum_{l=1}^{70}A_{kl}B_l \ .
\end{equation}
$A$ is a $61 \times 70$ matrix of rational functions of the kinematic variables and $\eps{}$. This matrix is implicitly defined in~\cite{Chicherin:2020oor}.
Furthermore, $B_l$ and $I_i$ will be the target of IBP reductions. In terms of the master integrals chosen by \fire, the IBP identities are given by
\begin{align}
    I_i &= \sum_{j=1}^{61}M_{ij}L_j \ , \\
    B_l &= \sum_{j=1}^{61}U_{lj}L_j \ ,
\end{align}
where $M_{ij}$ and $U_{lj}$ are $3 \times 61$ and $70 \times 61$ rectangular matrices, respectively, elements of which are rational functions in the kinematic variables and $\eps{}$.

We can then express $I_i$ in terms of $J_j^{(aux)}$ as
\begin{equation}\label{eq:ItoAux}
    I_i = \sum_{j=1}^{61}\sum_{k=1}^{61}M_{ij}\left(A N\right)^{-1}_{jk}J^{(aux)}_k \ .
\end{equation}
The matrix $\left(A N\right)$ is $61 \times 61$ and nonsigular, and the relation between $I_i$ and $J_j^{(aux)}$ is purely rational. We perform the required IBP reductions over a large prime number field, and utilize analytic reconstruction methods to compute~\eqref{eq:ItoAux} analytically. We then recover~\eqref{eq:ItoCan} by analytically composing the result with $R^{-1}$, and substitute the solutions of the finite integrals in terms of pentagon functions to obtain \cref{eq:pf1}.